\PassOptionsToPackage{unicode}{hyperref}
\PassOptionsToPackage{hyphens}{url}
\PassOptionsToPackage{dvipsnames,svgnames,x11names}{xcolor}
\documentclass[10pt,twocolumn,letterpaper]{article}
\usepackage{xcolor}
\usepackage[top=0.9in,bottom=0.8in,left=0.75in,right=0.75in,columnsep=0.3in]{geometry}
\usepackage{amsmath,amssymb}
\usepackage{iftex}
\ifPDFTeX
  \usepackage[T1]{fontenc}
  \usepackage[utf8]{inputenc}
  \usepackage{textcomp} % provide euro and other symbols
\else % if luatex or xetex
  \usepackage{unicode-math} % this also loads fontspec
  \defaultfontfeatures{Scale=MatchLowercase}
  \defaultfontfeatures[\rmfamily]{Ligatures=TeX,Scale=1}
\fi
\usepackage{lmodern}
\ifPDFTeX\else
\fi
\IfFileExists{upquote.sty}{\usepackage{upquote}}{}
\IfFileExists{microtype.sty}{% use microtype if available
  \usepackage[]{microtype}
  \UseMicrotypeSet[protrusion]{basicmath} % disable protrusion for tt fonts
}{}
\usepackage{setspace}
\makeatletter
\@ifundefined{KOMAClassName}{% if non-KOMA class
  \IfFileExists{parskip.sty}{%
    \usepackage{parskip}
  }{% else
    \setlength{\parindent}{0pt}
    \setlength{\parskip}{6pt plus 2pt minus 1pt}}
}{% if KOMA class
  \KOMAoptions{parskip=half}}
\makeatother
\makeatletter
\ifx\paragraph\undefined\else
  \let\oldparagraph\paragraph
  \renewcommand{\paragraph}{
    \@ifstar
      \xxxParagraphStar
      \xxxParagraphNoStar
  }
  \newcommand{\xxxParagraphStar}[1]{\oldparagraph*{#1}\mbox{}}
  \newcommand{\xxxParagraphNoStar}[1]{\oldparagraph{#1}\mbox{}}
\fi
\ifx\subparagraph\undefined\else
  \let\oldsubparagraph\subparagraph
  \renewcommand{\subparagraph}{
    \@ifstar
      \xxxSubParagraphStar
      \xxxSubParagraphNoStar
  }
  \newcommand{\xxxSubParagraphStar}[1]{\oldsubparagraph*{#1}\mbox{}}
  \newcommand{\xxxSubParagraphNoStar}[1]{\oldsubparagraph{#1}\mbox{}}
\fi
\makeatother

\usepackage{longtable,booktabs,array}
\usepackage{calc} % for calculating minipage widths
\usepackage{etoolbox}
\makeatletter
\patchcmd\longtable{\par}{\if@noskipsec\mbox{}\fi\par}{}{}
\makeatother
\IfFileExists{footnotehyper.sty}{\usepackage{footnotehyper}}{\usepackage{footnote}}
\makesavenoteenv{longtable}
\usepackage{graphicx}
\makeatletter
\newsavebox\pandoc@box
\newcommand*\pandocbounded[1]{% scales image to fit in text height/width
  \sbox\pandoc@box{#1}%
  \Gscale@div\@tempa{\textheight}{\dimexpr\ht\pandoc@box+\dp\pandoc@box\relax}%
  \Gscale@div\@tempb{\linewidth}{\wd\pandoc@box}%
  \ifdim\@tempb\p@<\@tempa\p@\let\@tempa\@tempb\fi% select the smaller of both
  \ifdim\@tempa\p@<\p@\scalebox{\@tempa}{\usebox\pandoc@box}%
  \else\usebox{\pandoc@box}%
  \fi%
}
\def\fps@figure{htbp}
\makeatother

\NewDocumentCommand\citeproctext{}{}

\makeatletter
 \let\@cite@ofmt\@firstofone
 \def\@biblabel#1{}
 \def\@cite#1#2{{#1\if@tempswa , #2\fi}}
\makeatother
\newlength{\cslhangindent}
\newlength{\csllabelwidth}
\newenvironment{CSLReferences}[2] % #1 hanging-indent, #2 entry-spacing
 {\begin{list}{}{%
  \setlength{\itemindent}{0pt}
  \setlength{\leftmargin}{0pt}
  \setlength{\parsep}{0pt}
  \ifodd #1
   \setlength{\leftmargin}{\cslhangindent}
   \setlength{\itemindent}{-1\cslhangindent}
  \fi
  \setlength{\itemsep}{#2\baselineskip}}}
 {\end{list}}
\usepackage{calc}

\newcommand{\CSLLeftMargin}[1]{\parbox[t]{\csllabelwidth}{\strut#1\strut}}
\newcommand{\CSLRightInline}[1]{\parbox[t]{\linewidth - \csllabelwidth}{\strut#1\strut}}

\usepackage{newunicodechar}
\newunicodechar{≥}{\ensuremath{\geq}}
\newunicodechar{≤}{\ensuremath{\leq}}
\newunicodechar{χ}{\ensuremath{\chi}}
\newunicodechar{∼}{\ensuremath{\sim}}
\newunicodechar{𝜎}{\ensuremath{\sigma}}
\usepackage{authblk}

\makeatletter
\@ifpackageloaded{caption}{}{\usepackage{caption}}
\AtBeginDocument{%
\ifdefined\contentsname
  \renewcommand*\contentsname{Table of contents}
\else
  \newcommand\contentsname{Table of contents}
\fi
\ifdefined\listfigurename
  \renewcommand*\listfigurename{List of Figures}
\else
  \newcommand\listfigurename{List of Figures}
\fi
\ifdefined\listtablename
  \renewcommand*\listtablename{List of Tables}
\else
  \newcommand\listtablename{List of Tables}
\fi
\ifdefined\figurename
  \renewcommand*\figurename{Figure}
\else
  \newcommand\figurename{Figure}
\fi
\ifdefined\tablename
  \renewcommand*\tablename{Table}
\else
  \newcommand\tablename{Table}
\fi
}
\@ifpackageloaded{float}{}{\usepackage{float}}
\floatstyle{ruled}
\@ifundefined{c@chapter}{\newfloat{codelisting}{h}{lop}}{\newfloat{codelisting}{h}{lop}[chapter]}
\floatname{codelisting}{Listing}

\makeatother
\makeatletter
\@ifpackageloaded{caption}{}{\usepackage{caption}}
\@ifpackageloaded{subcaption}{}{\usepackage{subcaption}}
\makeatother
\usepackage{bookmark}
\IfFileExists{xurl.sty}{\usepackage{xurl}}{} % add URL line breaks if available
\hypersetup{
  pdftitle={Faster alone vs.~further together: The speed and impact of researchers' responses to societal crises},
  pdfauthor={Nicholas A. Coles; Joao Francisco Goes Braga Takayanagi; Stephen M. Fiore; Lingfei Wu},
  pdfkeywords={team science, crisis
response, speed, impact, scientometrics},
  colorlinks=true,
  linkcolor=black,
  filecolor=black,
  citecolor=black,
  urlcolor=black,
  pdfcreator={LaTeX via pandoc}}

\usepackage[T1]{fontenc}
\usepackage{tgtermes}
\usepackage{newunicodechar}
\newunicodechar{Δ}{\ensuremath{\Delta}}
\newunicodechar{≈}{\ensuremath{\approx}}
\newunicodechar{−}{\ensuremath{-}}
\newunicodechar{×}{\ensuremath{\times}}
\newunicodechar{±}{\ensuremath{\pm}}
\newunicodechar{α}{\ensuremath{\alpha}}
\newunicodechar{β}{\ensuremath{\beta}}
\newunicodechar{τ}{\ensuremath{\tau}}
\usepackage{lato}

\usepackage{microtype}
\usepackage{fancyhdr}
\usepackage{titlesec}
\usepackage{enumitem}
\definecolor{hdrgrey}{gray}{0.45}
\titleformat{\section}{\sffamily\bfseries\large}{}{0pt}{}
\titleformat{\subsection}{\sffamily\bfseries\normalsize}{}{0pt}{}
\titleformat{\subsubsection}{\sffamily\itshape\normalsize}{}{0pt}{}
\titlespacing*{\section}{0pt}{1.4ex plus .3ex}{0.8ex}
\titlespacing*{\subsection}{0pt}{1.1ex plus .2ex}{0.5ex}
\titlespacing*{\subsubsection}{0pt}{0.9ex plus .2ex}{0.3ex}
\renewcommand{\headrulewidth}{0.4pt}
\renewcommand{\headrule}{\hbox to\headwidth{\color{hdrgrey}\leaders\hrule height \headrulewidth\hfill}}
\fancypagestyle{plain}{\fancyhf{}\renewcommand{\headrulewidth}{0pt}\fancyfoot[C]{\sffamily\footnotesize\color{hdrgrey}\thepage}}
\AtBeginDocument{\setlength{\parindent}{1.2em}\setlength{\parskip}{0pt}\setlength{\emergencystretch}{2em}\raggedbottom}
\title{Faster alone vs.~further together: The speed and impact of
researchers' responses to societal crises}

\begin{document}
\twocolumn[{%
\thispagestyle{plain}
\noindent{\sffamily\bfseries\small Article}\hfill{\sffamily\small\color{hdrgrey}Unpublished manuscript}\\[-0.6ex]
\rule{\linewidth}{1.2pt}\\[2.2ex]
{\sffamily\bfseries\fontsize{19}{23}\selectfont Faster alone vs.~further together: The speed and impact of researchers' responses to societal crises\par}
\vspace{2ex}
{\sffamily\normalsize Nicholas A. Coles$^{1,*}$, Joao Francisco Goes Braga Takayanagi$^{1}$, Stephen M. Fiore$^{2}$, Lingfei Wu$^{3}$\par}
\vspace{1.2ex}
{\sffamily\footnotesize\color{hdrgrey}$^{1}$Department of Psychology, University of Florida, Gainesville, USA. $^{2}$Institute of Simulation and Training, University of Central Florida, Orlando, USA. $^{3}$School of Computing and Information, University of Pittsburgh, Pittsburgh, USA. $^{*}$Corresponding author: \href{mailto:ncoles@ufl.edu}{ncoles@ufl.edu}\par}
\vspace{2.2ex}
\noindent\begin{minipage}{\linewidth}\setlength{\parindent}{0pt}\normalsize
Throughout the 21st century, society encountered numerous crises that
demanded quick and widespread action. In the present work, we documented
researchers' responses to 25 of these crises, analyzing 1,537,284
published responses from 3,262,152 researchers spanning 225 countries
and territories. An analysis of publication speed and citation patterns
yielded support for a proverbial claim: ``to go fast, go alone; to go
far, go together''. Solo authors were quicker to publish responses to
societal crises -- but their responses received less attention in
scholarly, news, and policy venues. Growth towards large-scale
collaboration was associated with increased impact; However, this team
size advantage was characterized by diminishing returns and
significantly varied across sampled crises. Furthermore, pluralistic
machine learning modeling highlights the importance of considering
additional context, such as government effectiveness and the expertise
of team leaders. Taken together, results facilitate evidence-based
conversations regarding how researchers optimize speed and impact in the
midst of societal crises.
\end{minipage}
\vspace{1.5ex}\noindent{\color{hdrgrey}\rule{\linewidth}{0.4pt}}\vspace{2ex}
}]

\setstretch{1}
Throughout history, human societies have encountered and navigated
crises that demanded quick, urgent, and widespread action. For example,
in the first quarter of the 21st century, society witnessed some of the
deadliest terrorist attacks in modern history, a global pandemic, and
the rapid release of AI technologies. For decades, researchers have
worked to document how people respond to these societal crises, helping
separate proverbial myths from reality. For example, contrary to common
belief, research suggests that people tend to respond to crises
\emph{not} with panic -- but with order, solidarity, and
cooperation\textsuperscript{1,2}.

Curiously, the actions of researchers themselves are often overlooked in
crisis response research -- leaving an incomplete narrative. For
example, to understand society's response to the September 11 terrorism
attacks, it is important to consider that a leading academic society
endorsed psychologists' participation in detainee
interrogation\textsuperscript{3}. To understand society's response to
the COVID-19 pandemic\textsuperscript{4}, it is important to consider
the record-setting speed in which researchers developed mRNA
vaccines\textsuperscript{5}. Last, to understand how society responds to
the shock of AI technologies, we argue it will once again be important
to consider the role of researchers -- such as the ultra-massive teams
that developed GPT-4\textsuperscript{6}, DeepSeek V3\textsuperscript{7},
and other technologies at the center of an ongoing ``AI
race''\textsuperscript{8}. In the present work, we document researchers'
responses to 25 of these societal crises, analyzing a total of 1,537,284
published responses from 3,262,152 researchers spanning 225 countries
and territories (Figure 1).

Proverbial claims about collaboration suggests ``to go fast, go alone;
to go far, go together''. This belief is also often echoed in the
scientific community. Proponents of team science often suggest that
collaboration is a way to ``go further,'' as it (a) facilitates role
specialization, (b) increases total effort dedicated to a problem, and
(c) creates conditions believed to facilitate recombinant growth, wisdom
of the crowd, and/or collective intelligence\textsuperscript{9--15}.
These benefits, however, are often suggested to come at a cost that may
be particularly noteworthy during crisis: speed. Indeed, proponents of
team science also often emphasize the time needed to (a) assemble teams,
(b) negotiate responsibilities, (c) navigate different institutional
policies, (d) manage disagreements, and so on\textsuperscript{16--19}.
Working alone is presumed to be quicker, with simpler coordination,
fewer bureaucratic constraints, and greater decision-making agility.

In the present work, we test proverbial claims at scale, examining
1,537,284 separate published responses from a total of 3,262,152
researchers. As methodological tests of speed, we borrowed from the
sociological notion of \emph{focusing events}\textsuperscript{20}. More
specifically, we sampled societal crises that (a) entered mainstream
awareness relatively suddenly, and (b) received relatively novel names
(e.g., ``Hurricane Katrina''; Figure 2, Panel A). This allowed us to
estimate response speed by comparing (a) the approximate date that most
people became aware of each crisis, to (b) the publication dates of
papers explicitly mentioning relevant keywords (retrieved via
\emph{OpenAlex}\textsuperscript{21}). To index the \emph{impact} of
researchers' published responses, we considered the number of citations
their work received in scholarly, news, and policy venues (tracked by
\emph{OpenAlex}\textsuperscript{21} and
\emph{Altmetric}\textsuperscript{22}). Last, to establish comparative
benchmarks, we compared team size to 12 other predictors often
considered at the individual (e.g., leader prestige), group (e.g.,
number of involved institutions), and societal level (e.g., political
stability). Here, we leveraged a pluralistic set of machine learning
models to derive robust estimates of the extent to which each variable
predicts the speed and impact of researchers' responses to
crises\textsuperscript{23}.

Expanding upon insights from seminal field studies\textsuperscript{1,2},
the present work provides a relatively comprehensive assessment of
researchers' published responses to an extreme phenomenon: societal
crises. Theoretically, this tests potentially extreme boundary
conditions of a robust finding in research on team science: work from
larger teams often achieves more impact (e.g., in scholarly articles,
patents, and policy venues), often at the cost of other potentially
desirable outcomes (e.g., innovativeness)\textsuperscript{24--29}. By
sampling a relatively large number of events, we further estimate the
\emph{reliability} of such relationships across a wide variety of crises
(e.g., the H1N1 pandemic, Brexit, and the announcement of an impending
``metaverse''). Taken together, our work not only lends itself to a more
complete narrative of how society responds to crisis, but also to
evidence-based conversations regarding how to optimize the speed and
impact of researchers' responses to these crises.

\section{Results}\label{results}

Similar to what was documented decades ago with the general population,
our results illustrate that researchers also often respond to crisis
with order, solidarity, and cooperation\textsuperscript{1,2}. In total,
we uncovered 1,537,284 published responses to 25 societal crises --
representing the collective response of 3,262,152 researchers from 225
countries and territories (Figure 1).

In many instances, we found that researchers do ``go alone''. Averaged
across crises, 49.1\% of published responses were attributed to a single
author (\emph{SD} = 22.0\% across crises). The other 50.9\% of the time,
researchers ``go together,'' with teams ranging from 2 to 100+
co-authors. Next, we describe the implications this decision has for the
speed and impact of their published responses.

\subsection{When responding to crises, solo authorship is faster but
less
impactful}\label{when-responding-to-crises-solo-authorship-is-faster-but-less-impactful}

Proverbial claims about collaboration suggests ``to go fast, go alone;
to go far, go together''. Consistent with such notions, researchers were
significantly quicker to publish responses as solo authors vs.~teams (Δ
= 100 days; Mann-Whitney \emph{W} = 1.88 × 10\textsuperscript{11},
\emph{p} \textless{} .001, rank-biserial \emph{r} = .18). Intriguingly,
however, the link between the speed and \emph{size} of teams was best
characterized as weakly quadratic -- as opposed to linear (Δ\emph{BIC} =
4195) or logarithmic (Δ\emph{BIC} = 1135; Figure 2, Panel B). More
specifically, speed most sharply decreased as researchers moved from (a)
solo to small team efforts (Δ = 96 days; Mann-Whitney \emph{W} = 1.14 ×
10\textsuperscript{11}, \emph{p} \textless{} .001, rank-biserial
\emph{r} = .18), vs.~(b) small to medium team efforts (Δ = 12 days;
Mann-Whitney \emph{W} = 1.09 × 10\textsuperscript{11}, \emph{p}
\textless{} .001, rank-biserial \emph{r} = .02). Speed then slightly
recovered as researchers moved towards their largest team efforts (see
Table S1 for additional descriptives).

Despite reductions in speed, team science was more impactful, in terms
of tracked mentions in scholarly, news, and policy venues (Figure 2,
Panel B). For example, consider a crisis-relevant paper published by a
solo author who ``went alone''. On average, these solo-authored papers
received 6.08 mentions in scholarly articles (\emph{SD} = 53.58), 0.34
mentions in the news (\emph{SD} = 6.45), and 0.03 mentions in policy
documents (\emph{SD} = 0.47). Comparatively, papers published by small
teams of 2 - 4 authors received (1) 149\% more mentions in scholarly
articles (\emph{M} = 15.11, \emph{SD} = 77.00; rank-biserial \emph{r} =
.29), (2) 57\% more mentions in the news (\emph{M} = 0.53, \emph{SD} =
9.75; rank-biserial \emph{r} = .02), and (3) 76\% more mentions in
policy documents (\emph{M} = 0.06, \emph{SD} = 0.77; rank-biserial
\emph{r} = .01; all Mann-Whitney \emph{p} \textless{} .001). As an even
more extreme example, papers published by medium-sized teams of 5 - 9
authors received (1) 311\% more mentions in scholarly articles (\emph{M}
= 24.95, \emph{SD} = 128.07; rank-biserial \emph{r} = .49), (2) 275\%
more mentions in the news (\emph{M} = 1.27, \emph{SD} = 16.52;
rank-biserial \emph{r} = .07), and (3) 181\% more mentions in policy
documents (\emph{M} = 0.10, \emph{SD} = 0.94; rank-biserial \emph{r} =
.02; all Mann-Whitney \emph{p} \textless{} .001).

\subsection{Movement towards large-scale collaboration yields
diminishing returns in
impact}\label{movement-towards-large-scale-collaboration-yields-diminishing-returns-in-impact}

Past field studies have documented a sociological tendency for people to
mass organize in response to crisis\textsuperscript{1}. Our results
illustrate a similar dynamic in science, not only in terms of the number
of teams but also their \emph{size}. Of 1,537,284 total published
responses to crises, we observed 122,353 publications (8.0\%) with ten
to over a hundred authors.

Observing a large number and range of big teams helps us better
understand their reliability and limits. Here, we once again replicate
patterns from classic field studies on crisis response: collaboration
growth yields diminishing returns\textsuperscript{30}. More
specifically, links between the size and measured impact of papers were
best characterized as logarithmic, as opposed to linear (Δ\emph{BIC} ≥
6,639; Wilcoxon signed-rank \emph{p} \textless{} .001) or quadratic
(Δ\emph{BIC} ≥ 5,203; Wilcoxon signed-rank \emph{p} \textless{} .001).
In other words, growth in team size was associated with diminishing
returns in measured impact. Notably, however, the rate of diminishing
returns significantly varied across crises; Including crisis-specific
random slopes consistently improved model fit (all Δ\emph{BIC} ≥ 8,063;
likelihood-ratio tests, all χ²(1) ≥ 8,091, \emph{p} \textless{} .001).
These results provide initial evidence of the importance of context:
teams of similar size respond more impactfully to some crises than
others.

\subsection{Size in context: Integrating predictors of quick and
impactful crisis responses in
science}\label{size-in-context-integrating-predictors-of-quick-and-impactful-crisis-responses-in-science}

Not surprisingly, many claims about the role of collaboration -- both
proverbial and scientific -- fail to consider size in context. For
example, perhaps it is difficult for people to ``go fast'' in the face
of political dysfunction or ``go far'' with relatively inexperienced
leaders. To illustrate the utility of considering team size alongside 12
other predictors measured at the individual, group, and societal level,
we deployed a pluralistic set of machine learning architectures (e.g.,
XGBoost, neural nets, random forests). When trained to consider the full
set of variables -- vs.~team size alone -- models were significantly
more accurate at predicting response speed (Δ\emph{R²} = .38) and
mentions in scholarly (Δ\emph{R²} = .39), news (Δ\emph{R²} = .09), and
policy venues (Δ\emph{R²} = .07) in a hold-out test dataset (all
Wilcoxon signed-rank \emph{p} \textless{} .001).

Considering team size along a dozen other variables also allows us to
compare how useful they are for predicting speed and impact. For
example, despite attracting significant attention from lay people and
researchers alike, permutation variable importance scoring indicates
that team size is \emph{not} the strongest sampled predictor of speed
and impact. Team size was the (a) 11th strongest sampled predictor of
response speed (model reliance \emph{M} = 0.16, \emph{SD} = 0.10), (b)
3rd strongest predictor of scholarly citations (\emph{M} = 0.15,
\emph{SD} = 0.05), (c) 2nd strongest predictor of news mentions
(\emph{M} = 0.75, \emph{SD} = 0.20), and (d) 2nd strongest predictor of
policy citations (\emph{M} = 0.61, \emph{SD} = 0.28; Figure 3). For
response speed, the strongest sampled predictors were expert-derived
ratings of the corresponding author's government -- most notably its
effectiveness, voice and accountability, and control of corruption (see
Figure 3). For mentions in scholarly, news, and policy venues, the
strongest sampled predictor was the \emph{h}-index of the corresponding
author.

\section{Discussion}\label{discussion}

Arguably every human society has had to navigate crises, whether they be
natural, technological, social, or of some other nature. Across
generations, these cumulative experiences have helped inspire folk
beliefs about how humans respond to crises --- some more accurate than
others. For instance, past research helped debunk beliefs that people
respond to crises with panic\textsuperscript{1,2}. Instead, it seems
that people respond with a surprising amount of order, solidarity, and
cooperation. Our results strengthen such assertions in the sense that we
show ≈1.6 million instances where researchers organized and published
responses to societal crises, approximately half of which involved a
\emph{team} of co-authors.

More centrally, we evaluated a popular but largely untested folk belief
about collaboration: ``to go fast, go alone; to go far, go together''.
Here, we found more support for conventional wisdom. When considering
publication dates and mentions across academic, news, and policy venues,
results suggest that researchers do indeed publish faster alone and more
impactfully together. However, results also point to oft-neglected
nuance: larger teams yield diminishing returns in impact. Furthermore,
context matters: team sizes that were impact-maximizing in some crises
(e.g., COVID-19) were less successful in others (e.g., responses to
September 11 terrorism attacks).

To further highlight the role of context, we considered the predictive
validity of 12 additional candidate variables. Results help illustrate
that a combination of individual-level, group-level, \emph{and}
society-level context is needed to understand effective crises
responses\textsuperscript{24}. For example, the prestige of the
corresponding author and the number of co-authors were consistently
strong predictors of measured impact, as was the number of papers a team
ultimately cites. For speed, conversely, the strongest predictors were
expert-derived ratings of facets of government effectiveness. In other
words, while it is generally the case that researchers publish faster
alone and more impactfully together, maximizing the speed and impact of
researchers' responses to crises will require considering context from
multiple angles.

In terms of limitations, we navigated conceptual challenges regarding
the operational definition of a crisis, a crisis response, and the
impact of those responses. We similarly wrestled with mechanistic
uncertainty --- e.g., uncertainty regarding whether large teams produce
vs.~flock to better ideas. Below, we elaborate on these challenges and
their implications for research on crises.

First, we encountered considerable differences in how societal crises
are defined. Scholars across different research traditions have
emphasized different characteristics, such as suddenness, urgency,
social disruption, and public perception\textsuperscript{31--33}. We did
not seek to resolve these long-standing debates. Instead, we sought a
working definition that permitted relatively clean tests of response
speed. Ultimately, we borrowed most heavily from the notion of sudden,
attention-grabbing \emph{focusing events}\textsuperscript{20}. Due to
their suddenness, these events allowed us to operationalize response
speed in a literal fashion: the number of days that passed since people
typically became aware of the development. However, in doing so, we
excluded a variety of societal crises, such as those that (a) developed
relatively gradually (e.g., climate change), (b) attracted less
attention (e.g., sand supply scarcity), and/or (c) have not received
historically unique names (e.g., recent waves of political
polarization). Crises such as these make it difficult to estimate
response speed, but nonetheless represent critical threats to modern day
societies. There is an urgent need to understand such developments more
thoroughly (e.g., via future analyses of self-reported timelines,
research repository commitment history, and/or grant activity reports).

We similarly faced difficulty defining what constitutes a crisis
response in science, as well as how to measure its impact. Here, our
operationalizations were largely shaped by data availability. We sought
large quantities of data because the phenomena we seek to understand are
rare; Societal crises represent an extreme disruption of norms, as do
the mass collaborations they often inspire. In the context of science,
we found the largest swaths of data in bibliometric repositories of
published scholarly papers, the researchers who authored them, and the
extent to which the work was subsequently discussed across multiple
communities. However, a drawback of this approach is that it misses
other ways that researchers respond to crises (e.g., via community
service) and seek to improve the world (e.g., increasing societal
well-being vs.~increasing citation counts). There are obvious challenges
to documenting those types of responses and quantifying those broader
forms of impact. Nonetheless, doing so is essential in order to develop
a comprehensive understanding of how researchers contribute to responses
to societal crises.

A third set of limitations involves mechanistic uncertainty. Put simply,
we do not know \emph{why} we observe correlations between the size,
speed, and impact of crisis responses in science. It is possible there
is a simple causal story, wherein simply increasing the size of one's
team leads to higher-quality insights. However, an equally plausible
story is that the most promising ideas better attract sustained
attention from larger groups of researchers. Similarly, it's possible
that large teams have stronger social capital instead of stronger ideas
--- elevating the impact of otherwise non-superior research. Evaluating
the mechanisms underlying such links is a oft-discussed goal in research
on teams\textsuperscript{27--29,34,35}, a goal we encourage future
researchers to address via methods that better facilitate causal
inference (e.g., field experiments, naturalistic experiments, etc.).

Until these limitations are better addressed, we caution against a
de-prioritization of solo responses or a fetishization of large-scale
responses. Effective crisis response likely requires a balance of solo
and large team effort from researchers, a mixture of traditionally and
non-traditionally tracked activities, and an eye towards outcomes rarely
measured or discussed in hiring and promotion.

To conclude, human societies have always had to navigate periods of calm
and crises --- a continuously repeating cycle. In the present work, we
take stock of the role of researchers in the aftermath of societal
crises, weighing in on a proverbial claim about collaboration: ``to go
fast, go alone; to go far, go together''. Across 1,537,284 published
responses from 3,262,152 researchers spanning 225 countries and
territories, we find strong support for this proverbial claim.
Researchers do indeed publish faster alone and more impactfully
together. However, growth towards large-scale collaboration yields
diminishing returns that vary across crises. Furthermore, a pluralistic
set of machine learning models converge on the importance of considering
additional context, such as government effectiveness or the expertise of
team leaders. In the future, we emphatically encourage stakeholders to
consider these insights -- as well as the potential interconnectedness
of different crisis responses, different types of activity and impact,
and the more conceptually challenging crises we did not sample in the
present work. After all, if society encounters crises we cannot
ultimately overcome, there will arguably be no institution of science
left to document and understand such developments.

\section{Materials and methods}\label{materials-and-methods}

\subsection{Ethics}\label{ethics}

To examine researchers' published responses to crisis events, we
interoperated secondary data on (1) research articles, authors, and
institutions (via \emph{OpenAlex}\textsuperscript{21}), (2) tracked
mentions in scholarly, news, and policy documents (via \emph{Altmetric
Explorer}\textsuperscript{22}), and (3) country-level governance
indicators (via WorldBank's \emph{Worldwide Governance
Indicators}\textsuperscript{36}). Because the analysis focused on
secondary data, the University of Florida Institutional Review Board
declared this work exempt (Protocol \#ET00042723). Anthropic Claude (3.7
Sonnet - Fable 5) was used to generate focusing event suggestions, as
well as to review all analysis code for accuracy, accessibility, and
reproducibility issues.

\subsection{Sampling crisis events}\label{sampling-crisis-events}

The sampling plan was designed to identify crises that provide
relatively clean methodological tests of response speed. To do so, we
borrowed from the sociological notion of sudden, attention-grabbing,
\emph{focusing events}\textsuperscript{20}. More specifically, we
identified societal crises that (a) entered mainstream awareness
relatively suddenly, and (b) received relatively novel names. This
allowed us to estimate response speed by simply comparing (a) the
approximate date that each crisis entered mainstream awareness, to (b)
the publication dates of papers explicitly mentioning relevant keywords
(retrieved via \emph{OpenAlex}\textsuperscript{21}). For example, there
are virtually no mentions of the phrase ``ChatGPT'' in the research
literature before 2022. However, in the three-year aftermath of its
release, researchers have mentioned it in 111,646 research publications
(Figure 2, Panel A).

To identify candidate focusing events, we considered (a) events
frequently discussed in crisis response research (e.g., the Deepwater
Horizon oil spill), (b) recommendations from foundational large language
models (e.g.~Claude 3.7 Sonnet), and (c) our own lived experiences. From
this list, we conducted initial bibliometric searches to ensure that the
crisis had relatively unique identifying keywords and large growth
around the approximate date of the focusing event. From this list, we
chose 25, seeking a \emph{variety} of different societal crises (Figure
1), including ones often characterized as natural (e.g., Hurricane
Katrina), sociopolitical (e.g., \#MeToo), technological (e.g., ChatGPT),
and economic realms (e.g., the 2008 Global Financial Crisis). This was
done to promote more generalizable conclusions about links between the
size, speed, and impact of researchers' responses to crises.

\subsection{Identifying published
responses}\label{identifying-published-responses}

Researchers respond to societal crises in many ways, including advising
policymakers, providing media commentary, and engaging the public
directly. Here, we focus on one of their most thoroughly tracked
contributions: publications. For each sampled crisis, we first
identified keywords or phrases that researchers used to explicitly
signal relevance. Second, we specified the approximate date that each
sampled crisis entered mainstream awareness. Third, we used
\emph{OpenAlex} to pull papers containing the relevant keywords in their
title, abstract, or (if available) main text.

As an example of this workflow, consider the COVID-19 pandemic. This
crisis entered mainstream awareness near late December 2019, when
initial reports of the virus emerged from Wuhan, China. We sampled
papers that made an explicit connection to this crisis by using keywords
like ``COVID-19'', ``SARS-CoV-2'', and ``novel coronavirus''. See
\emph{Table S3} for full list of keyword strings.

In total, we identified 1,610,388 papers containing keywords linked to
our 25 sampled crises within the three years before and after each
event. Consistent with the notion of focusing events, only 2.9\% of this
keyword use occurred in the 3 years \emph{preceding} the approximate
date that each crisis entered mainstream awareness (Figure 2, Panel A).
Our analysis focuses on the 97.1\% of papers published in the 3 years
\emph{following} these events. After removing non-articles and records
without tracked authorship meta-data, this produced a final sample of
1,537,284 papers.

\subsection{Operationalizing response speed and
impact}\label{operationalizing-response-speed-and-impact}

For each paper, we operationalized response speed as the number of days
between (a) their publication date and (b) the approximate date each
crisis entered mainstream awareness. For impact, we considered multiple
indicators: the number of times each paper was mentioned in scholarly
works (\emph{N} = 27,898,094 mentions examined), the news (\emph{N} =
1,462,597 mentions), and policy documents (\emph{N} = 115,423 mentions).
Scholarly citation counts were pulled from \emph{OpenAlex} article
meta-data\textsuperscript{21}. News and policy document citation counts
were pulled from \emph{Altmetric Explorer}\textsuperscript{22}. Because
a Digital Object Identifier is required to match a paper to an
\emph{Altmetric Explorer} record, analyses of response speed and
scholarly citations draw on all 1,537,284 papers, whereas analyses of
news and policy mentions draw on the 1,370,810 papers carrying one (see
\emph{Table S1} and \emph{Table S2} for the sample sizes behind each
estimate). To reduce sensitivity to outliers, speed and impact indices
were converted to percentiles when fitting regression models. Because
crises vary in the overall amount of attention they receive, percentiles
were calculated separately for each event (i.e., standardized). (See
\emph{Figure S1} for similar results without this standardization.) For
ease of interpretation, percentiles are always reported in a positive
direction, wherein higher values indicate faster and more impactful
publication.

\subsection{Predictors of quick and impactful crisis responses in
science}\label{predictors-of-quick-and-impactful-crisis-responses-in-science}

The primary analyses focused on evaluating proverbial claims about links
between the size, speed, and impact of researchers' published responses
to crises. Team size was extracted from \emph{OpenAlex} meta-data. For
most analyses, team size was treated as a continuous variable. However,
for ease-of-interpretation, we also report common (but
arbitrary\textsuperscript{28,37}) categorizations: papers published by a
(a) solo author, (b) small teams (2 - 4 authors), (c) medium teams (5 -
9 authors), and (d) big teams (10+ authors).

The secondary analyses focused on considering team size alongside three
other sets of predictors. The first captures other characteristics of
teams and their work, such as the unique number of institutions and
countries involved, the primary discipline, and the volume of literature
cited. The second set of predictors focused on the person who leads the
work -- more specifically the corresponding author's past academic
success (\emph{h}-index) and type of home institution (education,
healthcare, government, company, non-profit, archival). These predictors
were all extracted from OpenAlex meta-data\textsuperscript{21}.

The third set of predictors considered the corresponding author's
broader societal context. More specifically, we connected their
affiliation data to the World Bank's Worldwide Governance Indicators,
matched by year of publication (up to last recorded year: 2022) and
country\textsuperscript{36}. These indicators contain expert ratings of
six dimensions, with higher values indicating normatively better
governance: (1) voice and accountability (e.g., freedom of expression),
(2) political stability (e.g., absence of violent destabilization), (3)
government effectiveness (e.g., quality of public services), (4)
regulatory quality (e.g., ability to promote private sector
development), (5) rule of law (e.g., adherence to social rules), and (6)
control of corruption (e.g., non-privatization of public power). Because
the World Bank series is biennial before 2002, intervening years were
filled by within-country linear interpolation.

\subsection{Analytic strategy}\label{analytic-strategy}

\subsubsection{Primary analysis}\label{primary-analysis}

The primary analysis focused on evaluating the extent to which links
between the size and effectiveness of crisis responses are linear,
logarithmic, or quadratic. Conceptually, this represents constant,
diminishing, and curvilinear returns, respectively. For example, rather
than leading to constant linear gains, it is possible that increases in
team size lead to diminishing returns. This would suggest that
researchers navigate a plateau, wherein they must make increasingly
larger investments to increase the speed and impact of their crisis
responses. Alternatively, increases in team size may exhibit a
curvilinear ``Goldilocks zone'': increases in team size may lead to
improved performance up to a point, after which further growth becomes
\emph{detrimental}. Such relationships are captured via linear
(\(\beta_{1}X_{i}\)), logarithmic (\(\beta_{1}ln(X_{i})\)), and
quadratic regression (\(\beta_{1}X_{i} + \beta_{2}X_{i}^2\)),
respectively.

Linear, logarithmic, and quadratic relationships between team size and
each outcome were modeled separately using the lme4 mixed-effect
modeling package\textsuperscript{38} (v. 2.0.1) in R (v. 4.5.3). To
accommodate the fact that papers published on similar crises are
non-independent, we included random intercepts for each sampled crisis:
\(\beta_{0} + u_{0,j}\), wherein \(u_{0,j} ∼ N(0, 𝜎_0^2)\). To ensure
that relatively uncommon large team efforts (e.g., 20 co-authors)
received the same priority as relatively common small team efforts
(e.g., 2 co-authors) during model fitting, we applied inverse frequency
weights. (See \emph{Figure S1} for similar patterns from raw data
plots.) When comparing fit, we favored a model (e.g., logarithmic over
linear) when two criteria were both met: (a) it reduced the Bayesian
Information Criterion by more than 5 (Δ\emph{BIC} \textgreater{} 5), and
(b) a one-sided paired Wilcoxon signed-rank test indicated that its
distribution of weighted absolute residuals was significantly lower
(using \emph{p} \textless{} .05 threshold). Mann-Whitney comparisons of
team-size categories were two-sided. No adjustments were made for
multiple comparisons. Given the unusually large size of our sample, we
focus more on interpreting patterns than statistical significance.

To evaluate the extent to which the best-fitting relationship varied
across sampled crises, random slopes were added. For example, when a
logarithmic model best fit the data, random slopes would be included for
the logarithmic term: \((\beta_{1} + u_{1, j})ln(X_{ij})\), wherein
\(u_{1,j} ∼ N(0, 𝜎_1^2)\). The present work mostly focuses on describing
the results of these models. Nonetheless, for convenience, we also
report comparisons of team size categories; e.g., papers published by
small (2 - 4 authors) vs.~medium teams (5 - 9 authors). In these
scenarios, we used Mann-Whitney tests to compare group distributions
(using \emph{p} \textless{} .05 threshold).

\subsubsection{Secondary analysis}\label{secondary-analysis}

A complete understanding of how researchers respond to crises may
ultimately require considering team size alongside other variables that
may confound or moderate its effects. To do so, we fit a pluralistic set
of machine learning algorithms containing a fuller set of potentially
relevant predictors. (For more information on predictor selection, see
\emph{Predictors of quick and impactful crisis responses in science}.)
To evaluate the utility of adding these additional predictors, we
compared models with vs.~without their inclusion. Because different
modeling approaches can produce meaningful variation in researchers'
conclusions\textsuperscript{39}, we used AutoGluon
1.5.0\textsuperscript{40} to train a variety of different models for
each outcome: gradient-boosted decision trees (LightGBM, CatBoost, and
XGBoost), randomized tree ensembles (random forests and extremely
randomized trees), and neural networks. When fitting models, we sought
authoritative inferences by using best-quality AutoGluon presets, with
dynamic stacking disabled and no additional stacking levels. Each model
was given up to 60 hours to train.

Prior to model training, we divided papers by crisis and then randomly
assigned 80\% into a training/validation set. The remaining 20\% was
held out as test data. Crucially, dividing papers by crisis before
splitting data helped ensure that their statistical patterns were
considered during both training and testing.

To estimate the importance of each predictor, we used a model-agnostic
permutation method\textsuperscript{23}. This methodology estimates the
importance of each predictor by randomly shuffling its values and then
re-testing the accuracy of each algorithm. When the algorithm relies on
the predictor, randomly shuffling its values diminishes performance. We
applied this permutation-based method to each predictor of each
top-fitting base model for each of the 4 outcomes. Importance was
estimated with five random shuffles per predictor on a random subsample
of 20,000 hold-out papers. To facilitate comparability across outcomes
and models, variable importance values were rescaled to a {[}0, 1{]}
range within each base model before being plotted in Figure 3.

\section{References}\label{references}
\small

\protect\phantomsection\label{refs}

\normalsize
\section{Data availability}\label{data-availability}

Processed data and code are available at
\url{https://github.com/Social-Science-Collaboratory/btmobile_sharing}
(release assets include the processed data). Data were sourced from
OpenAlex (CC0), Altmetric Explorer (licensed; cannot be redistributed),
and the World Bank Worldwide Governance Indicators (CC-BY 4.0).
Altmetric data cannot be openly shared because of its proprietary
nature. For users without licenses, we provide simulated copies of the
data. For users with licenses, we provide information on how to
replicate the data download.

\section{Code availability}\label{code-availability}

Code is available at
\url{https://github.com/Social-Science-Collaboratory/btmobile_sharing}.

\section{Author contributions}\label{author-contributions}

Conceptualization: N.A.C., L.W., J.F.G.B.T., S.M.F. Data curation:
N.A.C., J.F.G.B.T. Formal analysis: N.A.C., J.F.G.B.T. Funding
acquisition: N.A.C. Investigation: N.A.C., J.F.G.B.T. Methodology:
N.A.C., L.W., S.M.F., J.F.G.B.T. Project administration: N.A.C.,
J.F.G.B.T. Software: N.A.C., J.F.G.B.T. Supervision: L.W., S.M.F.,
N.A.C. Validation: J.F.G.B.T., N.A.C. Visualization: N.A.C., J.F.G.B.T.
Writing - original draft: N.A.C. Writing - review \& editing: N.A.C.,
J.F.G.B.T., L.W., S.M.F.

\section{Competing interests}\label{competing-interests}

The authors declare no competing interests.

\newpage{}

\section{Figures}\label{figures}

\begin{figure*}

\centering{

\includegraphics[width=1\linewidth,height=\textheight,keepaspectratio]{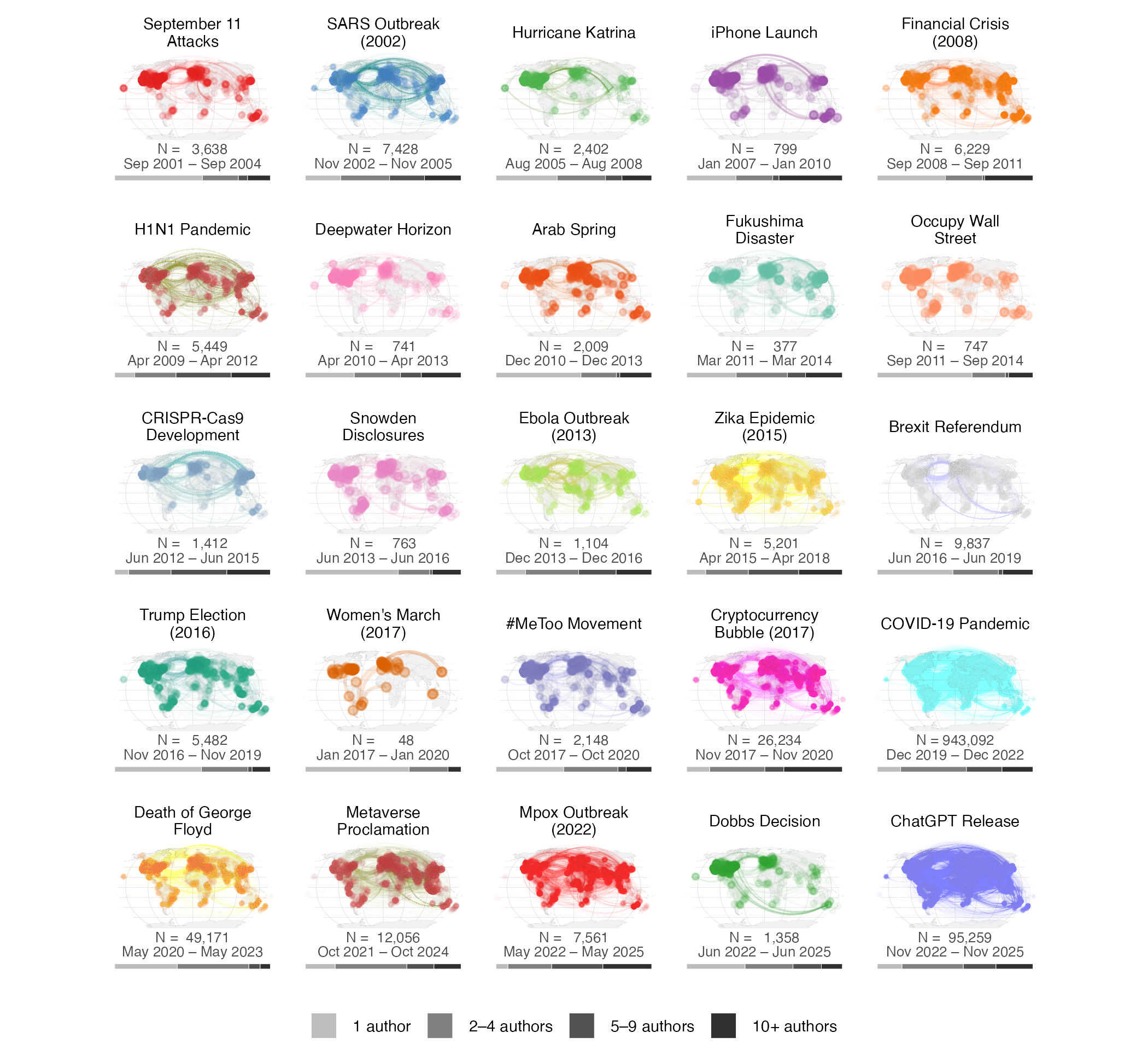}

}

\caption{\label{fig-map}\textbf{Overview of sample characteristics.} For
each of the 25 sampled crises, plotting panels include (a) the analyzed
date range, (b) the total number of uncovered records, and (c) the
proportion of papers published alone or with a small (2 - 4 authors),
medium (5 - 9 authors), or large (10+ authors) team, and (d) the
geographical distribution of authors, connected by co-authorship. For
legibility, the three most densely sampled crises (Death of George
Floyd, COVID-19, and ChatGPT) are drawn from a random subsample of
author institutions and co-authorship links.}

\end{figure*}%

\newpage{}

\begin{figure*}

\centering{

\includegraphics[width=1\linewidth,height=\textheight,keepaspectratio]{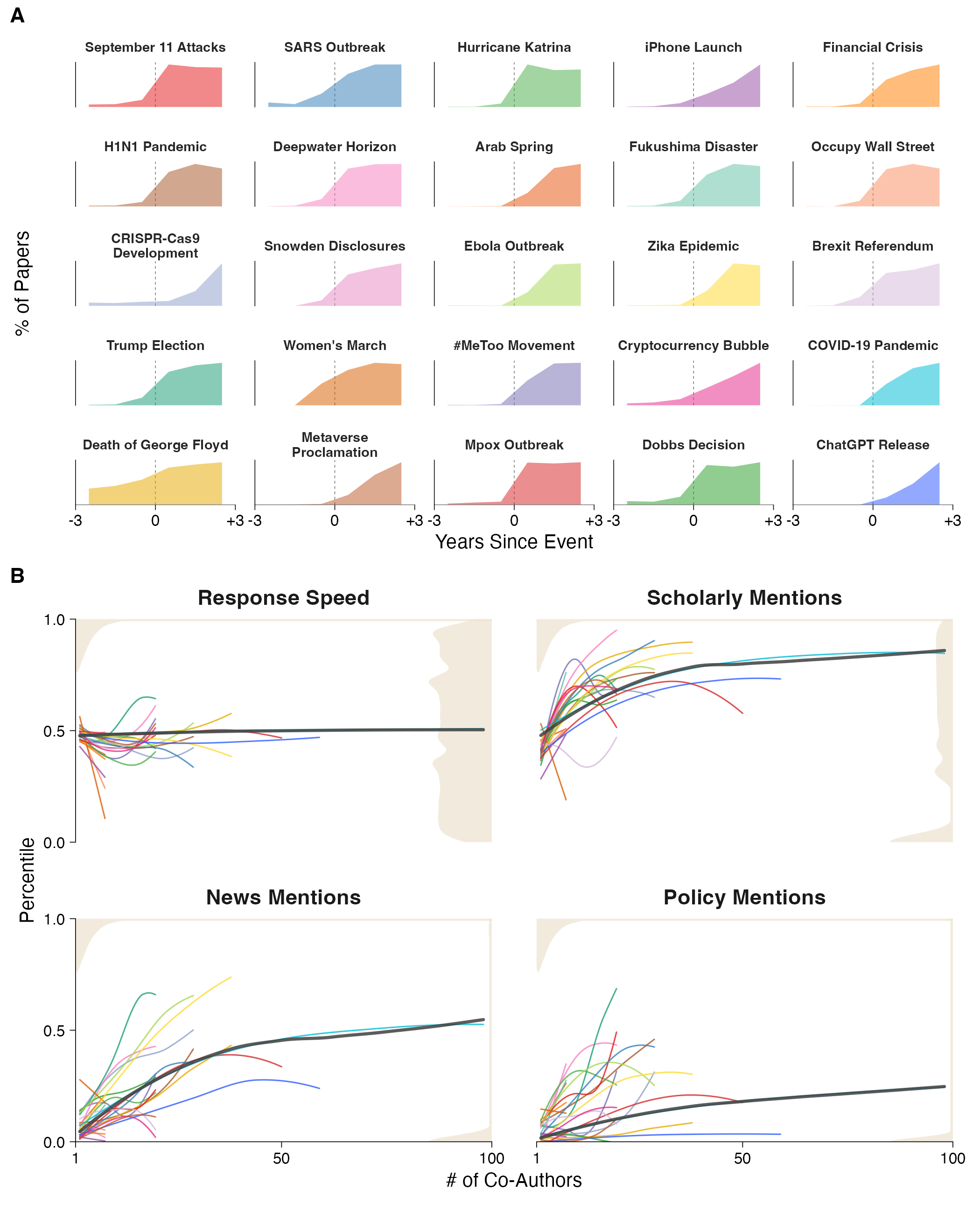}

}

\caption{\label{fig-area-impact}\textbf{Speed and impact of researchers'
responses to 25 societal crises.} Panel A: For each sampled crisis, the
proportion of keyword-linked papers published in the three years prior
and after the approximate date it entered mainstream awareness (dotted
line). Panel B: Smoothed links between the size (x-axis) and
effectiveness (y-axis) of published responses. For ease of
visualization, response speed and mentions in scholarly, news, and
policy venues are converted to percentiles and plotted separately.
Colored lines represent crisis-specific patterns, whereas the dark grey
line represents overall patterns. Marginal distributions of co-author
counts (top) and outcome percentiles (right) are shown in beige.}

\end{figure*}%

\newpage{}

\begin{figure*}

\centering{

\includegraphics[width=1\linewidth,height=\textheight,keepaspectratio]{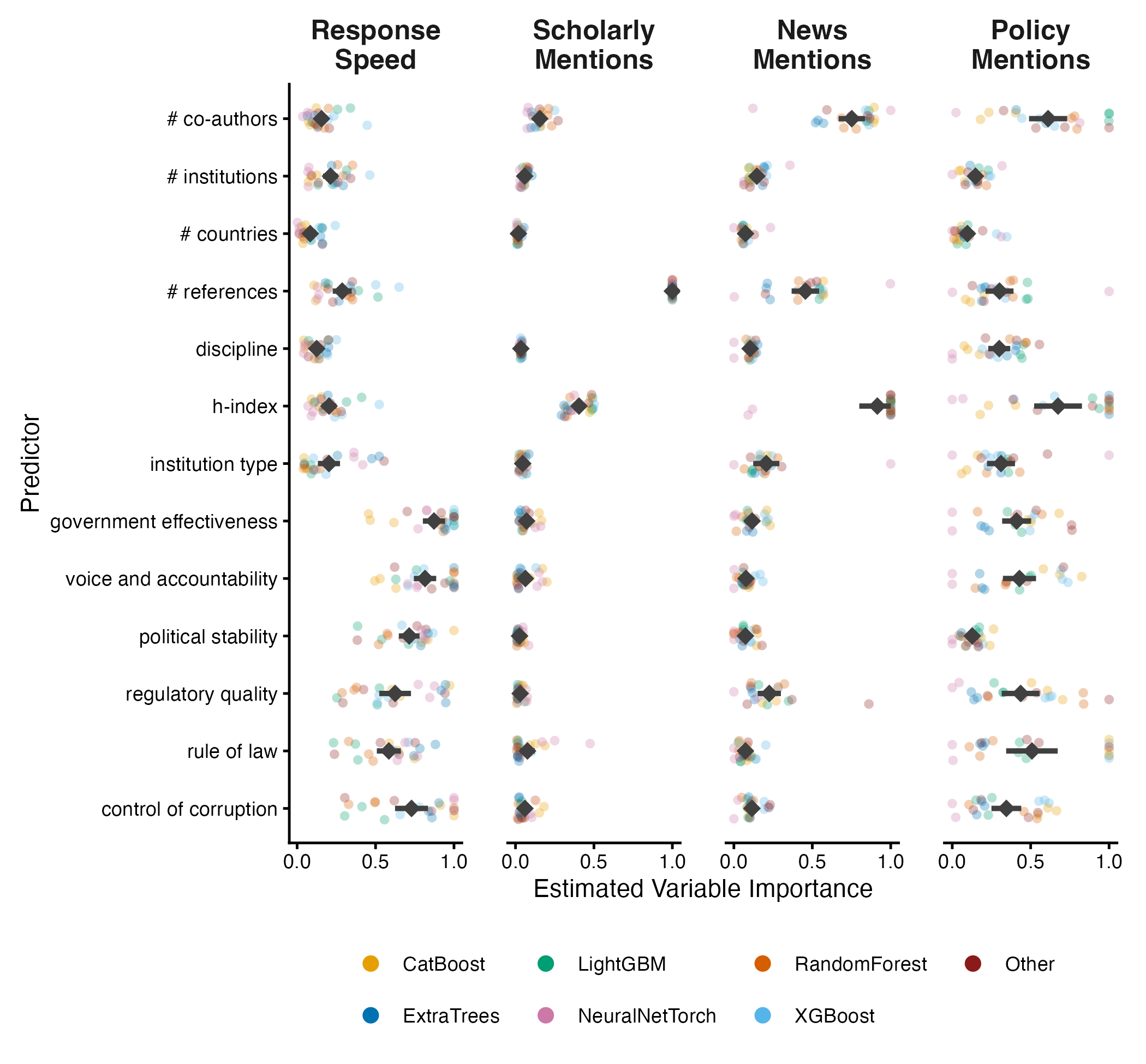}

}

\caption{\label{fig-varimp}\textbf{Predictors of quick and impactful
crisis responses.} Estimated permutation variable importance (x-axis) of
13 predictors (y-axis) for each of the four primary outcomes (plotted in
separate panels), from the secondary machine-learning analysis.
Importance is scaled to {[}0, 1{]} within each base model, denoted by
color. Points represent individual base models, diamonds the mean across
base models, and bars the 95\% confidence interval.}

\end{figure*}%

\end{document}